\documentclass[12pt]{article}

\usepackage{epstopdf,amsfonts,amsmath}
\usepackage{graphicx}
\usepackage{tikz}
\usepackage{stmaryrd}

\DeclareGraphicsExtensions{.eps}

\newcommand\encadremath[1]{\vbox{\hrule\hbox{\vrule\kern8pt
\vbox{\kern8pt \hbox{$\displaystyle #1$}\kern8pt}
\kern8pt\vrule}\hrule}}
\def\enca#1{\vbox{\hrule\hbox{
\vrule\kern8pt\vbox{\kern8pt \hbox{$\displaystyle #1$}
\kern8pt} \kern8pt\vrule}\hrule}}

\newcommand\figureframex[3]{
\begin{figure}[bth]
\hrule\hbox{\vrule\kern8pt
\vbox{\kern8pt \vbox{
\begin{center}
{\mbox{\epsfxsize=#1.truecm\epsfbox{#2}}}
\end{center}
\caption{#3}
}\kern8pt}
\kern8pt\vrule}\hrule
\end{figure}
}
\newcommand\figureframey[3]{
\begin{figure}[bth]
\hrule\hbox{\vrule\kern8pt
\vbox{\kern8pt \vbox{
\begin{center}
{\mbox{\epsfysize=#1.truecm\epsfbox{#2}}}
\end{center}
\caption{#3}
}\kern8pt}
\kern8pt\vrule}\hrule
\end{figure}
}

\makeatletter
\@addtoreset{equation}{section}
\makeatother
\newtheorem{theorem}{Theorem}[section]

\newtheorem{remark}{Remark}[section]
\newtheorem{proposition}{Proposition}[section]
\newtheorem{lemma}{Lemma}[section]
\newtheorem{corollary}{Corollary}[section]
\newtheorem{definition}{Definition}[section]
\def\br{\begin{remark}\rm\small}
\def\er{\end{remark}}
\def\bt{\begin{theorem}}
\def\et{\end{theorem}}
\def\bd{\begin{definition}}
\def\ed{\end{definition}}
\def\bp{\begin{proposition}}
\def\ep{\end{proposition}}
\def\bl{\begin{lemma}}
\def\el{\end{lemma}}
\def\bc{\begin{corollary}}
\def\ec{\end{corollary}}
\def\beaq{\begin{eqnarray}}
\def\eeaq{\end{eqnarray}}

\newcommand{\eq}[1]{eq.~(\ref{#1})}

\newcommand{\beq}{\begin{equation}}
\newcommand{\eeq}{\end{equation}}
\newcommand{\beqq}{\begin{equation}}
\newcommand{\eeqq}{\end{equation}}
\newcommand{\bea}{\begin{eqnarray}}
\newcommand{\eea}{\end{eqnarray}}
\newcommand{\beaa}{\begin{eqnarray*}}
\newcommand{\eeaa}{\end{eqnarray*}}

\newcommand{\Tr}{\operatorname{Tr}}
\newcommand{\e}{{\rm e}}

\newcommand{\Lieg}{{\mathfrak g}}
\newcommand{\Lieh}{{\mathfrak h}}
\newcommand{\LieU}{{\mathfrak U}}

\newcommand{\Adj}{\operatorname{Adj}}

\newcommand{\curve}{{\Sigma}}

\newcommand{\curveuniv}{{\tilde\curve}}

\newcommand{\Res}{\mathop{\,\rm Res\,}}
\usepackage[pdftex]{hyperref}
\hypersetup{colorlinks,urlcolor=magenta,citecolor=red,linkcolor=blue,filecolor=black}

\begin{document}

\hfill Preprint: IPhT-16/007, CRM-3352
\sloppy

\addtolength{\baselineskip}{0.20\baselineskip}
\begin{center}
\vspace{1cm}

{\Large \bf {Loop equations from differential systems}}

\vspace{1cm}

{Bertrand Eynard}$^{1,2}$,
{Rapha\"el Belliard}$^1$,
{Olivier Marchal}$^3$

\vspace{5mm}
$^1$\ Institut de physique th\'eorique, Universit\'e Paris Saclay, 
\\
CEA, CNRS, F-91191 Gif-sur-Yvette, France
\vspace{5mm}
\\
$^2$\ Centre de recherches math\'ematiques, Universit\'e de Montr\'eal, Canada
\vspace{5mm}
\\
$^3$\  Universit\'{e} de Lyon, CNRS UMR 5208, Universit\'{e} Jean Monnet,
\\
Institut Camille Jordan, France
\vspace{5mm}
\\
\end{center}

\vspace{1cm}
\begin{center}
{\bf Abstract}

To any differential system $d\Psi=\Phi\Psi$ where $\Psi$ belongs to a Lie group (a fiber of a principal bundle) and $\Phi$ is a Lie algebra $\Lieg$ valued 1-form on a Riemann surface $\curve$, is associated an infinite sequence of ``correlators" $W_n$ that are symmetric $n$-forms on $\curve^n$. 
The goal of this article is to prove that these correlators always satisfy "loop equations", the same equations satisfied by correlation functions in random matrix models, or the same equations as Virasoro or W-algebra constraints in CFT.

\end{center}
\begin{quote}

\end{quote}




\section{Introduction}

Given $\Lieg$ a reductive Lie algebra and $G=e^{\Lieg}$ its connected Lie group (think of $G=GL_r(\mathbb C)$ and $\Lieg=M_r(\mathbb C)$), we will consider the linear differential equation $\nabla \Psi=0$ satisfied by a flat section $\Psi$ in a principal $G-$bundle over a complex curve $\curve$, equipped with a connection $\nabla$. 
Locally this takes the form $d\Psi=\Phi\Psi$ where the ``Higgs field" $\Phi$ is locally a $\Lieg$ valued holomorphic 1-form.
To the data of a flat section $\Psi$, and a choice of a faithful representation $\rho$ of $\Lieg$, is associated \cite{BE09,BBE14,BE10} an infinite tower of ``correlators" called $W_n$ (definition recalled below). These correlators naturally appear in many contexts like Matrix Models, Conformal Field Theory (CFT) \cite{CER13}, some Painlev\'e equations \cite{P5,P2,IwakiMarchal}, or in Cohomological Field Theories \cite{BDY14}.

In the context of matrix models, ``loop equations" are an infinite set of algebraic relations satisfied by the $W_n$s. They are usually obtained by integration by parts and are also called ``Schwinger-Dyson" equations because they can also be derived from the invariance of an integral under changes of integration variable. The name ``loop equations" for Schwinger-Dyson equations of matrix models was coined by A.Migdal in \cite{Migdal:1984gj}, as these played a huge role in the quantum gravity matrix model activities in the 1990s \cite{DGZ}, and it was realized that loop equations were formally Virasoro or W-algebra constraints \cite{DGZ}.

However, loop equations can be generalized beyond the context of matrix models, just as a set of {\bf algebraic relationships among the $W_n$s}.

In \cite{BE09}, the authors derived loop equations in the case $\Lieg=\mathfrak{sl}_2(\mathbb C)$ on the Riemann sphere. However, the proof in \cite{BE09} involved an ``insertion operator", that was hard to define rigorously in all cases, and involved analysis (infinitesimal deformations). It was unsatisfactory because loop equations are algebraic statements, that cry for an algebraic proof.

Then in \cite{BBE14}, the authors found a purely algebraic derivation of a subset of loop equations (those with $n=0$ in the notations below), for $\Lieg=\mathfrak{sl}_r(\mathbb C)$.
In \cite{EynardRibault16} it was realized that the natural language is to work with a Lie algebra, and the authors found a completely general algebraic proof of loop equations, although in \cite{EynardRibault16} it was only restricted to Fuchsian systems on compact Riemann surfaces.

The purpose of this paper is to prove loop equations in an algebraic manner in a totally general case. Somehow this can serve as a lemma to be used in many applications.

A consequence of having loop equations, is that, if our differential system satisfies further nice properties (called ``topological type", see section \ref{secTT}), then we automatically have ``topological recursion" \cite{EOFg}.

To sum up, the goal of this article is to prove that correlators of local Hitchin systems always satisfy loop equations.

\section{Lie algebra Hitchin pair on a Riemann surface}

Let $\Lieg$ be a reductive Lie algebra \cite{bourbakiLie} (think of $\Lieg=\mathfrak gl_r(\mathbb C)=M_r(\mathbb C)$ the algebra of complex $r\times r$ matrices).
Let $\rho$ be a faithful representation of $\Lieg$ into the vector space of complex $r\times r$ matrices $M_r(\mathbb C)$, and define the invariant form of $\Lieg$ by
\beq
<a,b> = \Tr \rho(a) \rho(b)\stackrel{{\rm def}}{=:}  \Tr_\rho a b .
\eeq
Being invariant means $<[a,b],c>=<a,[b,c]>$ and $<g a g^{-1} , g b g^{-1}> = <a.b>$. On a reductive Lie algebra, there is no unique invariant form, our definition thus depends on a choice of a faithful representation $\rho$. If we would suppose $\Lieg$ to be semi-simple, then the invariant form would not depend on $\rho$ apart from a trivial multiplication by a non-zero constant. In other words, it would be the Killing form. However our general setting does not require semi--simplicity and therefore we do not assume it.

Let $\curve$ be a Riemann surface. $\curve$ may not be compact, it may have punctures, boundaries, high genus, etc. It does not matter since the loop equations proved in this article are local. Typically $\curve$ may be an open disc of $\mathbb C$.

Let $\mathcal E$ be a (possibly twisted\footnote{
The prime form initially defined by Fay in \cite{Fay} is not defined on $\curve$, only on the universal cover: it has monodromies. Fay also defined twisted prime forms, that have no monodromies, but that may have essential singularities and poles. Here, we may restrict our Riemann surface to a sub-domain that excludes those singularities.}) ``prime form" on $\curve\times \curve$, i.e. a $(-1/2,-1/2)$ form  that behaves on the diagonal like
\beq
\mathcal E(x,x') \sim \frac{x-x'}{\sqrt{dx dx'}} + O((x-x')^2),
\eeq
in any choice of local coordinates,
and has no other zeros on $\curve\times\curve$.
In particular, we do not require anti-symmetry, i.e. possibly $\mathcal E(x',x) \neq -\mathcal E(x,x')$. We also allow singularities away from the diagonal.
(see \cite{Fay} for a definition of prime forms on compact curves).
On the Riemann sphere or $\mathbb C$, one can just choose 
\beq\label{primeformonC}
 \mathcal E(x,x') = \frac{x-x'}{\sqrt{dx dx'}}.
\eeq

\bigskip

Let $(\mathcal P,\Phi)$ be a Hitchin pair \cite{Hitchin}, where $\mathcal P$ is a principal $G$-bundle over $\curve$, and $\nabla=d-\Phi$ a connection, where $\Phi$, called the Higgs field, is a $\Lieg$-valued holomorphic 1-form on $\curve$ (up to redefining $\curve$ by removing the singularities of $\Phi$, without loss of generality).
Let $\Psi$ be a locally flat section, i.e. satisfying $\nabla \Psi=0$, written locally as a differential system
\beq
d\Psi = \Phi\Psi.
\eeq
$\Psi$ is actually defined on a universal cover $\curveuniv$ of $\curve$.
Any two flat sections are related by a right multiplication:
\beq
\tilde \Psi(x) = \Psi(x) C
\,\,, \,\, C\in G \,\,{\rm independent \, of}\,x,
\eeq
where the choice of $C$ corresponds to a choice of initial condition at a point used to define the universal cover.

\section{Correlators}

To this connection $\nabla=d-\Phi$, a flat section $\Psi$, and a faithful representation $\rho$, we shall associate a tower of  ``correlators" $W_n$. They are used for example in matrix models \cite{BE09,DGZ}, in CFT \cite{CER13,ER14,EynardRibault16} or in cohomological field theories in \cite{BDY14}. Their definition, first introduced in \cite{BE09,BBE14,BookEynard} is recalled below.

\smallskip

We denote $\mathcal P_0$ the trivialized $\Lieg$-bundle with constant fiber $\curveuniv\times \Lieg \to \curveuniv$, with trivial connection $d$, i.e. whose flat sections are constant sections.

\beq
\begin{array}{ccl}
\mathcal P & & \mathcal P_0 =  pr^* \mathcal P = \curveuniv\times \Lieg \cr
p\downarrow & \overset{\pi}{\swarrow} & \downarrow \pi_0 \cr
\curve & \underset{pr}{\longleftarrow} & \curveuniv
\end{array}
\eeq

We denote $pr$ the projection $\curveuniv\to \curve$, and $\pi=\pi_0^*\, pr$ the projection $\mathcal P_0\to \curve$.

Throughout the rest of the paper, $X=\tilde x.E$ will denote a point in the total space of $\mathcal P_0$, in other words $\tilde x\in \curveuniv$ and $E\in \Lieg$, and $x=\pi(X)=pr(\tilde x)\in \curve$.

Besides, we denote $\Adj \mathcal P$ the adjoint bundle of $\mathcal P$, whose $\Lieg$ fiber over $x\in \curve$, is $\Lieg_x=T_{1_{G_x}} G_x$ the tangent space of the $G_x$ fiber of $\mathcal P$ at $x$ (with the same transition functions as $\mathcal P$), and equipped with the adjoint connection $d-\Adj_\Phi$. 

\bd
We introduce the bundle morphism $M: \mathcal P_0 \to \Adj \mathcal P$
defined by
\beq
M(\tilde x.E) :=   \Adj_{\Psi(\tilde x)}(E) = \Psi(\tilde x) E \Psi(\tilde x)^{-1}.
\eeq
It sends flat sections of $\mathcal P_0$ (i.e. constant $E$) into flat sections of the connection $d-\Adj_\Phi$ on the adjoint bundle. In other words we have locally, at constant $E$:
\beq
dM(X) = [\Phi(\pi(X)),M(X)].
\eeq

\ed
In case $\curve$ is not simply connected, its fundamental group is non--trivial. A nice property of the bundle map $M$ is that it descends to the quotient by a fundamental group action.
Indeed let $\underline{\pi}_1(\curve)\to\curve$ be the family of fundamental groups over $\curve$ (the fundamental groupoid).
After going around a loop $\gamma\in \pi_1(\curve,pr(\tilde x))$, $\Psi$ picks a monodromy $\Psi(\tilde x+\gamma)=\Psi(\tilde x)S_{\gamma}$, and thus $M(\tilde x+\gamma.E) = M(\tilde x.\Adj_{S_\gamma} (E)) = M(\tilde x.S_\gamma E S_\gamma^{-1})$. Consequently we introduce:
\bd
Let
\beq
\hat\curve = {\mathcal P_0} / {\underline{\pi}_1(\curve)}
\eeq
where the fiberwise quotient is relative to the $\underline{\pi}_1(\curve)$ action defined by  $\gamma.(\tilde x.E) = (\tilde x+\gamma).\Adj_{S_\gamma^{-1}}( E)$ for every $\gamma\in \pi_1(\curve,pr(\tilde x))$, i.e. we identify $\tilde x.E\equiv (\tilde x+\gamma).\Adj_{S_\gamma^{-1}}( E)$ in $\mathcal P_0$.

We see that $M$ can be pushed to the quotient, and using the same name $M$ for the pushforward to $\hat\curve$, we have $M\in Bun_{\Sigma}(\hat\Sigma, \Adj P)$, which means that $M$ maps 
$\hat\curve$  into $\Adj \mathcal P$. We also denote $\pi$ the projection from $\hat\curve$ to the base curve $\curve$:
\beq
\begin{array}{rll}
\hat\curve & \stackrel{M}{\hookrightarrow} & \Adj \mathcal P \cr
\pi &  \searrow & \downarrow \cr
& & \curve
\end{array}
\eeq
\ed

Remark  that changing the choice of flat section $\Psi\to \Psi C$ or changing the choice of universal cover and fundamental group (both depend on a choice of a base point on $\curve$), amounts to an isomorphism $\mathcal P_0 \to \Adj_C \mathcal P_0$ obtained by conjugation of each fiber by a constant group element $C$. Modulo such isomorphisms, $\hat\curve$ and the correlators $W_n$ to be defined below, will depend only on a connection $d-\Phi$, but not on a choice of local flat section $\Psi$.

\bd[Connected Correlators]

Let $\rho$ be a faithful representation of $\Lieg$, extended to the universal enveloping algebra $\LieU$ of $\Lieg$ \cite{bourbakiLie}.

Let $X=[\tilde x.E]$, and $X_i=[\tilde x_i.E_i]$ be some points of $\hat\curve$ (i.e. equivalence classes of $\curveuniv\times \Lieg$ modulo the $\underline{\pi}_1(\curve)$ action), with projections $x_i=\pi(X_i)$ all distinct on $\curve$, we define:
\beq
\hat W_1(X) = < M(X),\Phi(\pi(X)))> = \Tr_\rho \left(M(X)\Phi(\pi(X))\right) ,
\eeq
\beq
\hat W_2(X_1,X_2) = -\,\frac{< M(X_1), M(X_2)>}{\mathcal E(x_1,x_2) \mathcal E(x_2,x_1)} = -\,\frac{\Tr_\rho M(X_1) M(X_2)}{\mathcal E(x_1,x_2) \mathcal E(x_2,x_1)},
\eeq
and for $n\geq 3$, 
\beq
\hat W_n(X_1,\dots,X_n) = \sum_{\sigma \in \mathfrak S_n^{1-{\rm cycle}}} (-1)^{\sigma} \frac{\Tr_\rho M(X_1) M(X_{\sigma(1)}) M(X_{\sigma^2(1)}) \dots  M(X_{\sigma^{n-1}(1)})}{\mathcal E(x_1,x_{\sigma(1)}) \mathcal E(x_{\sigma(1)},x_{\sigma^2(1)})  \dots \mathcal E(x_{\sigma^{n-1}(1)},x_{1})}
\eeq
where the sum is over all permutations that have exactly one cycle (in particular with signature $(-1)^\sigma=(-1)^{n-1}$).

We recall that we have chosen $<a,b>=\Tr_\rho ab$, and we define $\Tr_\rho a_1 a_2 \dots a_n := \Tr \rho(a_1) \dots \rho(a_n) = \Tr \rho(a_1\otimes a_2 \otimes \dots \otimes a_n)$ in $\LieU$. 

\ed
$\hat W_1$ is a 1-form on $\hat\curve$, and $\hat W_n$ is a symmetric $n-$form on $\hat\curve^n$ (see \cite{EynardRibault16}). Then let us define the full correlators (so far we have defined the ``connected" correlators):

\bd[Correlators] We define the correlators by:
\beq
W_n(X_1,\dots,X_n)
= \sum_{\mu \vdash \{X_1,\dots,X_n\}}
\prod_{i=1}^{\ell(\mu)} \hat W_{|\mu_i|}(\mu_i)
\eeq
where we sum over all partitions of the set $\{X_1,\dots,X_n\}$ of $n$ points.
For example
\beq
W_1(X_1) = \hat W_1(X_1),
\eeq
\beq
W_2(X_1,X_2)
= \hat W_1(X_1) \hat W_1(X_2) + \hat W_2(X_1,X_2)
\eeq
\bea
W_3(X_1,X_2,X_3)
&=& \hat W_1(X_1) \hat W_1(X_2) \hat W_1(X_3) + \hat W_1(X_1) \hat W_2(X_2,X_3) \cr
&& + \hat W_1(X_2) \hat W_2(X_1,X_3) + \hat W_1(X_3) \hat W_2(X_1,X_2) \cr
&& + \hat W_3(X_1,X_2,X_3) 
\eea
and so on...

\ed

\subsubsection{CFT notation}

Very often we shall denote correlators as in the physics CFT notations with some Sugawara \cite{sugawara} bosonic $\Lieg$-currents\footnote{Often in the literature, the currents are written in a basis $e_1,\dots,e_{\dim \Lieg}$ of $\Lieg$, as vectors $\vec J(\tilde x) = (J_1(\tilde x),\dots,J_{\dim\Lieg}(\tilde x))$, with $J_k(\tilde x) = J(\tilde x.e_k)$.} $J(X_i)$:
\beq
W_n(X_1,\dots,X_n) =\left\langle J(X_1) \dots J(X_n) \,V_\Phi\right\rangle
\eeq
where $V_\Phi$ is a CFT operator depending on our choice of Higgs field, typically, if $\Phi$ is Fuchsian (only simple poles), then $V_\Phi$ is a product of vertex operators at the poles $p_i$ of $\Phi$ with charges ${\bf\alpha}_i=\Res_{p_i} \Phi$,  as $V_\Phi = \prod_{p_i={\rm poles}} V_{{\bf\alpha}_i}(p_i)$.
It is explained in \cite{EynardRibault16} why these are indeed Sugawara conformal blocks correlators: they satisfy OPEs and Ward identities of a $\Lieg$ Kac-Moody CFT at central charge $c={\rm rank}\,\Lieg$.
The relationship between CFT and differential systems is also observed for example in \cite{DoreyTateo, MS15}.

%
%

\subsection{Determinantal formulas}

Let us define the kernel:
\beq
K(\tilde x_1,\tilde x_2) = \frac{\Psi(\tilde x_1)^{-1} \Psi(\tilde x_2)}{\mathcal E(x_1,x_2)}
\eeq
where $x_i=pr(\tilde x_i)$,  the parallel transport kernel of the connection $d-\Phi$ (indeed $\mathcal E(\pi(x_1),\pi(x_2))\Psi(\tilde x_1) K(\tilde x_1,\tilde x_2) =  \Psi(\tilde x_2)$).
Let us define its ``normal ordered" version denoted (borrowed from CFT notations) by dots $:K:$, obtained by subtracting the pole when points are coinciding on the base
\beq
:K(\tilde x_1,\tilde x_2): = 
\left\{\begin{array}{l}
\frac{\Psi(\tilde x_1)^{-1} \Psi(\tilde x_2)}{\mathcal E(x_1,x_2)} \qquad {\rm if}\quad x_1\neq x_2 \cr
\cr
\Psi(\tilde x_1)^{-1} \Phi(x_1) \Psi(\tilde x_1) \qquad {\rm if}\quad x_1= x_2 \cr
\end{array}\right.
\eeq
$K(\tilde x_1,\tilde x_2)$ is a locally $(1/2,1/2)$ form on $\curveuniv\times\curveuniv$, 
taking values in $G_{x_1}\times G_{x_2}$ (the Lie group fibers over the points $\tilde x_1$ and $\tilde x_2$ of the principal bundle $\mathcal P$),
and with a simple pole at $x_1=x_2$. Its regularization at $x_1=x_2$ is a $\Lieg$-valued 1-form. We have
\bt
If $pr(\tilde x_1),\dots ,pr(\tilde x_n)$ are all distinct:
\beq
W_n(\tilde x_1.E_1,\dots,\tilde x_n.E_n) = \Tr \sum_{\sigma\in\mathfrak S_n} (-1)^\sigma \prod_i \rho(E_i) \rho(:K(\tilde x_i,\tilde x_{\sigma(i)}) :)
\eeq
which, by abuse of notation, we may denote as a determinant, whence the name ``determinantal formula":
\beq\label{defnormalorderWn}
W_n(\tilde x_1.E_1,\dots,\tilde x_n.E_n) = \Tr_\rho :\det E_i K(\tilde x_i,\tilde x_j) :
\eeq
here the determinant  means the sum over permutations of products of $E$s and $K$s taking values in $\LieU$, of which we finally take the trace in representation $\rho$.

\et

\section{Loop equations}

\subsection{Casimirs}

Let $e_1,\dots,e_{\dim\Lieg}$ be an arbitrary basis of $\Lieg$.
Since the invariant pairing $<a,b>=\Tr \rho(a)\rho(b)$ is the restriction to $\Lieg$ of the non-degenerate canonical pairing in $M_r(\mathbb C)$, and since we assume $\rho$ faithful, then $<,>$ is not degenerate on $\Lieg$, and therefore there exists a unique dual basis  $e^1,\dots,e^{\dim\Lieg}$ of $\Lieg$ such that 
\beq
\forall\, i,j\in \llbracket 1,\Lieg\rrbracket\,:\,\,  <e_i,e^j>=\delta_{i,j}.
\eeq

The enveloping algebra $\LieU$ of $\Lieg$ is defined as
\bd
\beq
\LieU = \left(\underset{k=0}{\overset{\infty}{\oplus}} \Lieg^{\otimes k}\right)/ <a\otimes b - b\otimes a - [a,b]>
\eeq

\ed

The Casimirs are elements of the center $Z(\LieU)$, they can be obtained as follows.

Let $\rho$ be a faithful representation of $\Lieg$ into $M_r(\mathbb C)$.
Let $v=\underset{i=1}{\overset{\dim\Lieg}{\sum}} v^i e_i\in \Lieg$. The characteristic polynomial of $\rho(v)$ is a symmetric polynomial of the coordinates $v^1,\dots,v^{\dim\Lieg}$, that can be  written
\beq
\text{det}_{\rho}(y-v) := \det(y {\rm Id}_r -\rho(v))  = \sum_{k=0}^r (-1)^k y^{r-k} \sum_{1\leq i_1,\dots,i_k\leq \dim\Lieg} C_k(i_1,\dots, i_k) v^{i_1} \dots v^{i_k} 
\eeq
Then we have the classical result (\cite{LieAlgebraGeneral}):
\bt
The Casimirs
\beq
C_k =\sum_{1\leq i_1,\dots,i_k\leq \dim\Lieg} C_k(i_1,\dots, i_k) e^{i_1} \otimes \dots \otimes e^{i_k}  \in \LieU
\eeq
are in the center of $\LieU$.
In fact the $C_k$s generate $Z(\LieU)$, but in general they are not algebraically independent.

\et

For example in a semi--simple Lie algebra the second Casimir is
\beq
C_2 = -\,\frac12  \sum_{i=1}^{\dim\Lieg} e_i\otimes e^i.
\eeq

\bt
The same Casimirs can be obtained with a basis of a Cartan subalgebra only.
Let $\Lieh$ a Cartan subalgebra of $\Lieg$, with an arbitrary basis $e_1,\dots,e_{\dim\Lieh}$ and $e^1,\dots,e^{\dim\Lieh}$ its dual basis $<e_i,e^j>=\delta_{i,j}$.
Let $v=\underset{i=1}{\overset{\dim\Lieh}{\sum}} v^i e_i\in \Lieh$. The characteristic polynomial of $\rho(v)$ is a symmetric polynomial of the coordinates $v^1,\dots,v^{\dim\Lieh}$, that we write
\beq
\text{det}_{\rho}(y-v) = \det(y {\rm Id}_r -\rho(v))  = \sum_{k=0}^r (-1)^k y^{r-k} \sum_{1\leq i_1,\dots,i_k\leq \dim\Lieh} \tilde C_k(i_1,\dots, i_k) v^{i_1} \dots v^{i_k} 
\eeq
Then the Casimirs are:
\beq
C_k = \sum_{1\leq i_1,\dots,i_k\leq \dim\Lieh} \tilde C_k(i_1,\dots, i_k) e^{i_1} \otimes \dots \otimes e^{i_k}  \in \LieU
\eeq
For example in a semi--simple Lie algebra
\beq
C_2 = -\,\frac{1}{2} \sum_{i=1}^{\dim\Lieh} e_i\otimes e^i.
\eeq

\et
This is a classical theorem in Lie algebras. In some sense it says that to compute the characteristic polynomial, we may choose a basis where $v$ is diagonal.

\subsection{$\mathcal W$ generators and Casimirs}

\bd Given $X_1,\dots,X_n$ points of $\hat\curve$ with distinct projections on $\curve$, and $\tilde x\in\curveuniv$, with $x=pr(\tilde x)$ distinct from the $\pi(X_i)$, we define:
\beq
W_{k;n}(C_k(x),X_1,\dots,X_n)
:= \sum_{1\leq i_1,\dots,i_k \leq \dim\Lieg} C_k(i_1,\dots,i_k) :W_{k+n}(\tilde x.e^{i_1},\dots,\tilde x.e^{i_k},X_1,\dots,X_n):
\eeq
It can be defined also using only the basis of a Cartan subalgebra $\Lieh$
\beq
W_{k;n}(C_k(x);X_1,\dots,X_n)
= \sum_{1\leq i_1,\dots,i_k \leq \dim\Lieh} \tilde C_k(i_1,\dots,i_k) :W_{k+n}(\tilde x.e^{i_1},\dots,\tilde x.e^{i_k},X_1,\dots,X_n):
\eeq
with the normal ordering defined in \eq{defnormalorderWn}.

It may seem that this definition depends on $\tilde x\in \curveuniv$ rather than $x\in\curve$, and also that it depends on a choice of basis of $\Lieg$ (or of $\Lieh$), but we shall prove below (loop equations) that it does not depend on a choice of a preimage $\tilde x\in pr^{-1}(x)$ of $x$, and is independent of the chosen basis of $\Lieg$ (resp. $\Lieh$).
\ed

As a Sugawara CFT notation we shall write it:
\beq
W_{k;n}(C_k(x);X_1,\dots,X_n) = \left\langle \mathcal W_k(x) J(X_1)\dots J(X_n) V_\Phi\right\rangle
\eeq
where $\mathcal W_k(x)$ is called the $k^{\rm th}$ W-algebra generator.
In particular for $k=2$ we denote $\mathcal W_2(x)=T(x)$  usually called the stress-energy tensor (up to a normalization).

\subsection{Loop equations}

We now reach the main theorem of this article.
This theorem can be interpreted as the Virasoro (or $\mathcal W$-algebra) constraints in a $\Lieg$--Kac--Moody CFT of central charge $c={\rm rank}\,\Lieg$.
\bt[Loop equations]
For any $n\geq 0$, and $X_1,\dots,X_n$ points of $\hat\curve$ with distinct projections $x_i=\pi(X_i)$, and $\tilde x\in \curveuniv$ also with distinct projection $x=pr(\tilde x)$ , we have
\small{\beq\label{loopmaineq}
\sum_{k=0}^r (-1)^k y^{r-k} W_{k;n}(C_k(x);X_1,\dots,X_n)
= [\epsilon_1\dots\epsilon_n]\, \text{det}_{\rho}\left( y-\left( \Phi(x) + \mathcal M_\epsilon(x;X_1,\dots,X_n)
\right)
\right)
\eeq}\normalsize{where} $y$ is a formal variable (a 1-form on $\curve$, the equality taking place in the determinant of the adjoint bundle), $[\epsilon_1 \dots \epsilon_n]$ is the notation indicating that we keep only the $\epsilon_1 \dots \epsilon_n$ coefficient of the Taylor expansion at $\epsilon_i\to 0$. Finally we have introduced the following symbol (that only makes sense in the representation $\rho$)
\bea
\mathcal M_\epsilon(x;X_1,\dots,X_n)
&=& \sum_{i=1}^n \epsilon_i \frac{M(X_i)}{\mathcal E(x,x_i) \mathcal E(x_i,x)} \cr
&& + \sum_{1\leq i\neq j\leq n} \epsilon_i \epsilon_j \frac{M(X_i)M(X_j)}{\mathcal E(x,x_i) \mathcal E(x_i,x_j) \mathcal E(x_j,x)} \cr
&& +  \sum_{k=3}^n \sum_{1\leq i_1\neq \dots \neq i_k\leq n } \epsilon_{i_1}\dots \epsilon_{i_k} \frac{M(X_{i_1})\dots M(X_{i_k}) }{\mathcal E(x,x_{i_1}) \mathcal E(x_{i_1},x_{i_2}) \dots  \mathcal E(x_{i_k},x)} \cr
\eea
The right hand side of \eqref{loopmaineq} is clearly an analytic function of $x\in\curve$ (rather than $\tilde x\in \curveuniv$) and is clearly independent of the chosen basis of $\Lieg$, which justifies the definition of the left hand side.
\et

{\bf Proof:}

Let us first consider the case $n=0$, already done in \cite{BBE14}.
By definition we have

\beaa
&& \sum_{k=0}^r (-1)^k y^{r-k} W_{k;0}(C_k(x)) \cr
&=& \sum_{k=0}^r (-1)^k y^{r-k} \sum_{1\leq i_1,\dots,i_k\leq \dim\Lieg} C_k(i_1,\dots,i_k) :W_{k}(\tilde x.e^{i_1},\dots,\tilde x.e^{i_k}):  \cr
&=& \sum_{k=0}^r (-1)^k y^{r-k} \sum_{1\leq i_1,\dots,i_k\leq \dim\Lieg} C_k(i_1,\dots,i_k)\sum_{\sigma\in\mathfrak{S}_k} (-1)^\sigma \Tr_\rho \prod_{j=1}^k\left( e^{\sigma(j)}:K(\tilde x,\tilde x):\right)\ \ \ \ \ \ \ \ \ \ \ \eeaa
but here, since all points have the same $x=pr(\tilde x)$, we have $:K(\tilde x,\tilde x) :=\Psi(\tilde x)^{-1} \Phi(x) \Psi(\tilde x)$, i.e.
\bea
&& \sum_{k=0}^r (-1)^k y^{r-k} W_{k;0}(C_k(x)) \cr
&=& \sum_{k=0}^r (-1)^k y^{r-k} \sum_{1\leq i_1,\dots,i_k\leq \dim\Lieg}  C_k(i_1,\dots,i_k)\sum_{\sigma\in\mathfrak{S}_k} (-1)^\sigma \Tr_\rho \prod_{j=1}^k\left( e^{\sigma(j)} \Psi(\tilde x)^{-1} \Phi(x) \Psi(\tilde x)\right) \cr
\eea
Using the cyclic property of the trace
\bea
&& \sum_{k=0}^r (-1)^k y^{r-k} W_{k;0}(C_k(x)) \cr
&=& \sum_{k=0}^r (-1)^k y^{r-k} \sum_{1\leq i_1,\dots,i_k\leq \dim\Lieg}  C_k(i_1,\dots,i_k)\sum_{\sigma\in\mathfrak{S}_k} (-1)^\sigma \Tr_\rho \prod_{j=1}^k\left( \Psi(x) e^{\sigma(j)} \Psi(x)^{-1} \Phi(x)\right)  \cr
\eea
Now, since the Casimirs are independent of which basis is chosen, change the basis $e_j\to \Psi(x) e_j \Psi(x)^{-1}$, and thus
\bea
&& \sum_{k=0}^r (-1)^k y^{r-k} W_{k;0}(C_k(x)) \cr
&=& \sum_{k=0}^r (-1)^k y^{r-k} \sum_{1\leq i_1,\dots,i_k\leq \dim\Lieg}  C_k(i_1,\dots,i_k)\sum_{\sigma\in\mathfrak{S}_k} (-1)^\sigma \Tr_\rho \prod_{j=1}^k  \left(e^{\sigma(j)}  \Phi(x)\right)  \cr
&=& \text{det}_{\rho}(y-\Phi(x))) 
\eea

The case $n\geq 1$ is similar. For any $k$
\bea
&& W_{k;n}(C_k(x);X_1,\dots,X_n) \cr
&=&  \sum_{1\leq i_1,\dots,i_k\leq \dim\Lieg} C_k(i_1,\dots,i_k) :W_{k+n}(\tilde x.e^{i_1},\dots,\tilde x.e^{i_k},X_1,\dots,X_n):  \cr
&=&  \sum_{1\leq i_1,\dots,i_k\leq \dim\Lieg}  C_k(i_1,\dots,i_k)\sum_{\sigma\in \mathfrak S_{k+n}} (-1)^\sigma \Tr_\rho \prod_{j=1}^{n+k} [\tilde{E}_{\sigma(j)}:K(\tilde x_{\sigma(j)},\tilde x_{\sigma(j+1)}) :] \cr
\eea
where now we sum over permutations of $k+n$ variables with for all 
$1\leq j\leq n$,  $X_j=[\tilde x_{j}.E_j]$ are representents of $X_j$,
while for all 
$n+1\leq n+j\leq n+k$ : $\tilde x_{n+j}=\tilde x$ and $E_{n+j}=e^j$.
If we define:
\bea
\tilde K_\varepsilon(\tilde x,\tilde x;X_1,\dots,X_n)
&=& :K(\tilde x,\tilde x) + \sum_{i=1}^n \varepsilon_i K(\tilde x,\tilde x_i)E_iK(\tilde x_i,\tilde x)  \cr
&& + \sum_{i\neq j=1}^{n} \varepsilon_i\varepsilon_j K(\tilde x,\tilde x_i)E_iK(\tilde x_i,\tilde x_j)E_j K(\tilde x_j,\tilde x) \cr
&& + \sum_{k=3}^n \sum_{i_1\neq \dots \neq  i_k=1}^{n} \varepsilon_{i_1} \dots \varepsilon_{i_k} K(\tilde x,\tilde x_{i_1})E_{i_1}K(\tilde x_{i_1},\tilde x_{i_2}) \dots E_{i_k} K(\tilde x_{i_k},\tilde x) :\cr
&=& \Psi(\tilde x)^{-1} \Big( \Phi(x) + \sum_{i=1}^n \varepsilon_i \frac{M(X_i)}{\mathcal{E}(x,x_i)\mathcal{E}(x_i,x)}   \cr
&& + \sum_{k=2}^n \sum_{i_1\neq \dots \neq  i_k=1}^{n} \varepsilon_{i_1} \dots \varepsilon_{i_k} \frac{M(X_{i_1})\dots M(X_{i_k})}{\mathcal{E}(x,x_{i_1}) \dots \mathcal{E}(x_{i_k},x)} \Big)\Psi(\tilde x) \cr
&=& \Psi(\tilde x)^{-1}(\Phi(x)+  \mathcal M_\varepsilon(x;X_1,\dots,X_n))\Psi(\tilde x)
\eea
The coefficient of $\epsilon_1\dots \epsilon_n$ in 
\beq
\sum_{\sigma\in \mathfrak S_{k}} (-1)^\sigma \Tr_\rho \prod_{j=1}^k\left( e^{\sigma(j)} \tilde K_\varepsilon(\tilde x,\tilde x;X_1,\dots,X_n) \right) 
\eeq
is a sum of products, where in each product each $M(X_i)$ appears exactly once, in all possible orders, and with products of $M([\tilde x , e^{i_k}])$ in between, thus it
exactly produces the sum over permutations of $k+n$ variables.
Therefore
\beaa
&& \sum_{k=0}^r (-1)^k y^{r-k} W_{k;n}(C_k(x);X_1,\dots,X_n) \cr
&=& [\varepsilon_1\dots\varepsilon_n] \sum_k (-1)^k y^{r-k} \sum_{1\leq i_1,\dots,i_k\leq \dim\Lieg} C_k(i_1,\dots,i_k) \cr
&& \qquad \qquad \underset{{\sigma\in \mathfrak S_{k}}}{\sum} (-1)^\sigma \Tr_\rho \underset{j=1}{\overset{k}{\prod}}\left( e^{\sigma(j)} \Psi(x)^{-1}  (\Phi(x)+\mathcal M_\varepsilon)\Psi(x)\right).
\eeaa
Using the same trick as in the case $n=0$ we can change the basis $e_j\to \Psi(x) e_j \Psi(x)^{-1}$ and we find:
\bea
\sum_{k=0}^r (-1)^k y^{r-k} W_{k;n}(C_k(x),X_1,\dots,X_n)=[\varepsilon_1\dots\varepsilon_n] \text{det}_{\rho}(y-(\Phi(x)+\mathcal M_\varepsilon(x;X_1,\dots,X_n))).\cr
\eea
The right hand side depends only on $x=pr (\tilde x)$ as announced, and is independent of a choice of basis of $\Lieg$.
This concludes the proof. $\square$

\subsubsection{Example}

Let us choose $\Lieg={\mathfrak gl}_r(\mathbb C)$. 
It is not semi-simple, it differs from $\mathfrak sl_r(\mathbb C)$ (which is semi--simple) by an Abelian $\mathbb C$, which shall factor out.
A Cartan subalgebra $\Lieh$ is the set $r\times r$ diagonal matrices.
Let us choose the following basis of $\Lieh$:
\beq
e_i=e^i ={\rm diag}(0,\dots,0,\overset{\overset{i}{\downarrow}}{1},0,\dots,0)
\eeq
the matrix whose only non-vanishing entry is at position $i$. 

Chose $\curve=\bar{\mathbb C}=\curveuniv$ to be the Riemann sphere, and the prime form as in \eqref{primeformonC}.
Let us define
\beq
\mathcal W_{i_1,\dots,i_n}(x_1,\dots,x_n) := W_n([x_1.e_{i_1}],\dots,[ x_n.e_{i_n}])
\eeq
viewed as a multivalued function of $x_1,\dots,x_n$ on an $r:1$ cover of $\curve$, with the index $i_k$ indicating that $x_k$ is in the $i_k^{\rm th}$ branch.

We have the "linear loop equation" (coefficient of $y^{r-1}$, i.e. the Trace, the Casimir $C_1$):
\beq
\sum_{i_1=1}^r \mathcal W_{i_1,\dots,i_n}(x_1,\dots,x_n) = \delta_{n,1}\Tr \Phi(x_1) + \delta_{n,2} \delta_{i_1,i_2} \frac{dx_1dx_2}{(x_1-x_2)^2}
\eeq
which is a holomorphic 1-form of $x\in \curve$.
Similarly, 
we have the "quadratic loop equation" (coefficient of $y^{r-2}$), i.e. 
the stress energy tensor:
\beq
\sum_{i_1<i_2} \mathcal W_{i_1,i_2}(x,x) = \frac12 \left( (\Tr\Phi(x))^2 - \Tr \Phi(x)^2\right)
\eeq
which is a holomorphic quadratic differential on $\curve$.
The stress--energy tensor times a current $J(x_3.e_{i_3})$ gives
\beq
\sum_{i_1<i_2} \mathcal W_{i_1,i_2,i_3}(x,x,x_3) =   \left(\Tr \Phi(x) \Tr M([x_3.e_{i_3}])-\Tr \Phi(x) M([x_3.e_{i_3}])\right) \,\,\frac{dx dx_3}{(x-x_3)(x_3-x)},
\eeq
and so on...

We thus recover the same loop equations as in \cite{BE09, BBE14}, i.e. the standard loop equations in Matrix Models.

\section{Asymptotic expansion and topological recursion}

A consequence of loop equations, is that it implies -- under good assumptions called ``topological type property" -- the topological recursion \cite{EOFg}.

We introduce a ``small" parameter $\hbar$, and consider a  1-parameter family of Higgs fields $\frac{1}{\hbar}\Phi(x,\hbar)$ for $\hbar \neq 0$. Thus, the family of differential equations for flat sections is locally
\beq
\hbar\, d \Psi(x,\hbar) = \Phi(x,\hbar)\,\Psi(x,\hbar).
\eeq

The purpose is to study asymptotically the $\hbar\to 0$ limit.

\subsection{Topological Type (TT) property}\label{secTT}
Following the work of \cite{BE09} and \cite{BBE14} we define the following topological type property:

\bd[Topological Type Property]\label{TT} 
The connection $\hbar\nabla = \hbar d-\Phi$ is said to be of ``{\bf topological type}" if and only if all the following conditions are met:

\begin{enumerate}

\item \underline{Asymptotic expansion}: 
There exists some simply connected open domain of $\curve$ (which allows to identify $\curveuniv=\curve$, and $\hat\curve=\mathcal P_0=\curve\times \Lieg$) and an Abelian subalgebra $\Lieh$ of $\Lieg$, in which the connected correlators $\hat W_n(X_1,\dots,X_n)$s with each $X_i\in \curve\times \Lieh$, have a Poincarr\'e asymptotic $\hbar$ expansion 
\beq
\hat W_n(X_1,\dots,X_n) = \frac{\delta_{n,1}}{\hbar} \hat W_1^{(0)}(X_1) + \sum_{k=0}^\infty \hbar^k \hat W^{(k)}_n(X_1,\dots,X_n),
\eeq
such that each $\hat W^{(k)}_n([x_1.E_1],\dots,[x_n.E_n])$ is, at fixed $E_i\in \Lieh$, an algebraic symmetric $n-$form of $x_1,\dots,x_n$.
In other words, there must exist  a (possibly nodal) Riemann surface $\mathcal{S}$ independent of $k$ and $n$, which is a ramified cover of $\curve$, such that the pullbacks, at fixed $E_i\in\Lieh$, of $\hat W^{(k)}_n([x_1.E_1],\dots,[x_n.E_n])$ to $\mathcal S^n$ are meromorphic symmetric $n$-forms.

\item \underline{Pole only at branchpoints}: For $(k,n)\notin\{(0,1),(0,2)\}$ and any $(E_1,\dots,E_n)\in \Lieh^n$, the connected correlation functions $\hat W_n^{(k)}([x_1.E_1],\dots,[x_n.E_n])$ pulled back to $\mathcal S$, may only have poles at the ramification points of $\mathcal{S}\to \curve$. 
In particular they cannot have singularities at nodal points of $\mathcal{S}$, or at the punctures, i.e. the pullbacks of singularities of $\Phi$.
Moreover $\hat W_2^{(0)}([x_1.E_1],[x_2.E_2])$ may only have a double pole along the diagonal of $\mathcal{S}\times \mathcal{S}$ of the form $\frac{dx_1dx_2 <E_1,E_2>}{(x_1-x_2)^2}$ but no other singularities. 

\item \underline{Parity}: Under the involution $\hbar \to -\hbar$:
\beq
\hat W_{n}|_{\hbar \hspace{+.1em} \mapsto - \hbar}([x_1.E_1],\dots,[x_n.E_n]) = (-1)^{n} \hat W_{n}([x_1.E_1],\dots,[x_n.E_n]).
\eeq

\item \underline{Leading order}: For all $n\geq 1$, the leading order of the series expansion in $\hbar$ of the correlation function $\hat W_n$ is at least of order $\hbar^{n-2}$. In other words:
\beq  
\forall n\geq 1,\,\, \forall \, 0\leq k\leq n-3 \,:\, \hat W_n^{(k)}([x_1.E_1],\dots,[x_n.E_n])=0 \eeqq    

\end{enumerate}

If the system has the topological property, we denote
\beq
\hat W_{g,n}(x_1.E_1,\dots,x_n.E_n) =\hat W^{(2g-2+n)}_n(x_1.E_1,\dots,x_n.E_n),
\eeq
and we have
\beq
\hat W_{n}(x_1.E_1,\dots,x_n.E_n) = \sum_{g=0}^\infty \hbar^{2g-2+n} \, \hat W_{g,n}(x_1.E_1,\dots,x_n.E_n) .
\eeq

\ed

All those properties are non-trivial, and there exists plenty of examples of $\Phi(x,\hbar)$ for which they are not satisfied.
Fortunately, there are also plenty of very interesting examples for which these conditions are satisfied. Let us recall certain \underline{sufficient conditions} under which these conditions may be satisfied.

\subsubsection{WKB expansion and condition $1$}

Condition $1$ can sometimes be obtained from asymptotic analysis, like it is done in large random matrices (where it is usually hard to prove).

Another method is to require condition $1$ as formal series. For example condition $1$ is always satisfied by formal WKB expansions.

Indeed, let introduce a ``small" parameter $\hbar$, and consider the Higgs fields $\frac{1}{\hbar}\Phi(x,\hbar)$, as a formal series of $\hbar$
\beq
\Phi(x,\hbar) = \sum_{k=0}^\infty \hbar^k\,\Phi^{(k)}(x).
\eeq
The formal family of differential equations for flat sections is locally
\beq
\hbar d \Psi(x,\hbar) = \Phi(x,\hbar)\,\Psi(x,\hbar).
\eeq

Let us choose once and for all a fixed Cartan subalgebra $\Lieh\subset \Lieg$ (think of $\Lieh$ as the set of diagonal matrices of $\Lieg=\mathfrak gl_r(\mathbb C)$).

The commutant of $\Phi^{(0)}(x)$ is generically a Cartan subalgebra, isomorphic to $\Lieh$, which means that 
$\Phi^{(0)}(x)$ can be ``diagonalized" as 
\beq
\Phi^{(0)}(x) = V(x) T'(x) V(x)^{-1} = \Adj_{V(x)} (T'(x))
\eeq
with $T'(x)$ a $\Lieh$-valued 1-form, and $V(x) \in G_x$ a group element. $V(x)$ and $T'(x)$ are defined up to a Weyl group action (permuting the eigenvalues) and invariant torus ($V(x)$ may be right-multiplied by an element of $e^{\Lieh}$).

In particular $T'(x)$ satisfies the algebraic equation
\beq
P(x,T'(x))=0 \qquad {\rm with}\quad P(x,y)=\text{det}_{\rho}(y-\Phi^{(0)}(x)) ,
\eeq
i.e. belongs to an algebraic plane curve $\mathcal{S}$ immersed in the total space of the cotangent bundle $T^*\curve$.
The immersion may or may not be an embedding, thus allowing nodal points for $\mathcal S$.

The characteristic polynomial $P(x,y)$ is called the \textbf{spectral curve} associated to the differential system. It defines a Riemann surface $\mathcal{S}$ with a projection to the base $x:\mathcal S \to \curve$, with some ramification points.

We define $T(x)$ a primitive of $T'(x)$ on the universal cover of $\curve$:
\beq
T(x) = \int_o^{x} T'(x')
\eeq
with $o$ an arbitrary base point. Changing $o$ or changing the integration path from $o$ to $x$ is just a shift of $T(x)$ by a constant, and will have no effect on what follows.

\bd
$\Psi(x,\hbar)$ is said to be  a formal WKB solution of $\hbar d\Psi=\Phi\Psi$, if and only if there exists a formal series of $\hbar$
\beq
\hat\Psi(x,\hbar) = {\rm Id} + \sum_{k=1}^\infty \hbar^k \hat\Psi^{(k)}(x),
\eeq
that satisfies to all powers of $\hbar$
\beq
\hbar\, d\hat\Psi = (V^{-1} \Phi V - \hbar V^{-1}dV) \hat\Psi - \hat\Psi T',
\eeq
i.e. such that
\beq
\Psi(x,\hbar) \sim V(x)\,\hat\Psi(x,\hbar) \e^{\frac{1}{\hbar}T(x)}
\eeq
is annihilated to all orders in $\hbar$, by $\hbar\nabla = \hbar d -\Phi(x,\hbar)$.

\ed

A formal WKB flat section $\Psi(x,\hbar)$ always exists, as can easily be seen by solving the equation $\hbar\, d\hat\Psi = (V^{-1} \Phi V - \hbar V^{-1}dV) \hat\Psi - \hat\Psi T'$ recursively in powers of $\hbar$. By doing so, we find $\hat\Psi^{(k+1)}(x)$ as an integral, and thus is not in general meromorphic on $\mathcal S$ since it may have monodromies. A sufficient condition (but not necessary) is that $\mathcal S$ is simply connected, i.e. if $\Phi^{(0)}(x)$ is meromorphic, we may require that the spectral curve $\mathcal S$ is a genus 0 curve.

From now on, let us consider a formal WKB solution $\Psi(x,\hbar)=V(x)\hat\Psi(x,\hbar) \,e^{\frac{1}{\hbar} T(x)}$.

Then, if we choose $E\in\Lieh$, we have
\beq
M(x.E) = \sum_{k\geq 0} \hbar^k M^{(k)}(x.E)
\eeq
where $M^{(k)}(x.E) = V(x) \underset{l=0}{\overset{k}{\sum}} \hat\Psi^{(l)}(x)E \hat\Psi^{(k-l)}(x)^{-1} V(x)^{-1}$.
In particular 
\beq
M^{(0)}(x.E) = V(x) E V(x)^{-1}.
\eeq
and thus if all $E_i$ are in $\Lieh$, $\hat W_n$ has a formal $\hbar$ expansion:
\beq
\hat W_n(x_1.E_1,\dots,x_n.E_n) = \frac{\delta_{n,1}}{\hbar} <T'(x_1),E_1> + \sum_{k=0}^\infty \hbar^k  \hat W^{(k)}_n(x_1.E_1,\dots,x_n.E_n).
\eeq
Thus WKB solutions satisfy condition $1$.

\subsubsection{Pole structure and condition $2$}

Generic WKB solutions obtained by recursively solving $\hbar\, d\hat\Psi = (V^{-1} \Phi V - \hbar V^{-1}dV) \hat\Psi - \hat\Psi T'$ in powers of $\hbar$, typically yield poles for the coefficients $\hat\Psi^{(k)}(x)$ whenever two eigenvalues of $\Phi^{(0)}(x)$ coincide, i.e. at the ramification points, but also at the nodal points.

Condition $2$ thus requires that poles at nodal points should cancel. This is a non-trivial condition, and many choices of $\Phi(x,\hbar)$ do not satisfy it.

In \cite{BE09, BBE14} it was realized that a sufficient condition for condition $2$, is that $\Phi(x,\hbar)$ is a Lax matrix, member of a time dependent family $\Phi(x,\hbar,t)$ that satisfies a Lax equation
\beq
\hbar \frac{\partial}{\partial t} \Phi(x,\hbar,t) = [\Phi(x,\hbar,t) , \mathcal R(x,\hbar,t) ] + \hbar \frac{\partial}{\partial x} \mathcal R(x,\hbar,t)
\eeq
with $\mathcal R(x,\hbar,t) = \underset{k=0}{\overset{\infty}{\sum}} \hbar^k \mathcal R^{(k)}(x,t)$ a formal series, whose spectral curve 
$$
\det( z-\mathcal R^{(0)}(x,t))=0,
$$
is a smooth embedding (no nodal point) in $T^*\curve$ (notice that $[\mathcal R^{(0)}(x,t),\Phi^{(0)}(x,t)]=0$, so that the two spectral curves have the same complex structure and same ramification points).

Under this assumption, the $\hat\Psi^{(k)}(x)$ can be found by recursively solving the ODE $\hbar \frac{\partial}{\partial t} \Psi = \mathcal R(x,\hbar,t) \Psi$, and it is then easy to see that there can be poles only at the branchpoints of the spectral curve of $\mathcal R^{(0)}$, i.e. only at the ramification points of $\mathcal S$, not at nodal points.

\subsubsection{Parity condition $3$}

A sufficient condition for the parity condition was found in \cite{BBE14}:

\begin{proposition}[Proposition $3.3$ of \cite{BBE14}]  
If there exists an invertible matrix $J$, independent of $x$, such that:
\beqq J^{-1} \Phi(x,\hbar)^tJ=\Phi(x,-\hbar)\eeqq
then the correlation functions $W_{n}$ satisfy: 
\beqq \forall \, n\geq 1\,:\, \hat W_{n}([x_1.E_1],\dots,[x_n.E_n],-\hbar) = (-1)^{n} \hat W_{n}([x_1.E_1],\dots,[x_n.E_n],\hbar)\eeqq
\end{proposition}

We do not know whether this condition is also a necessary one. We have not found any counter-example of a WKB+Lax system with parity property, not having a $J$ matrix.

\subsubsection{Leading order condition $4$}

Condition $4$ is often the most difficult to obtain. It is obvious that $\hat W_1$ is always $O(\hbar^{-1})$ and $\hat W_2$ is of order $O(\hbar^0)$ so that these cases are trivial. Moreover, from their definition all other $\hat W_n$s are at most of order $O(\hbar^0)$. If the parity is satisfied, then $\hat W_3$ must also be of order $O(\hbar)$ and thus is not a problem. But that $\hat W_4$ is of order $O(\hbar^2)$ rather than $O(\hbar^0)$ requires many non-trivial cancellations and the situation worsens when $n$ increases.

In \cite{BE09,BBE14}, was introduced axiomatically the notion of an ``insertion operator" mapping $\hat W_n\mapsto \hat W_{n+1}$ and being of order $\hbar$. We could prove the existence of such an insertion operator in very few cases like the $(p,2)$ minimal models in \cite{BookEynard}, and this was always non--trivial.
There is an incomplete proof for general $(p,q)$ minimal models in \cite{BBE14} and  Painlev\'{e} 5 in \cite{P5}, where only a subset of the requirements of an insertion operator were verified in \cite{BBE14}, it seems that the missing verifications could be done as in \cite{BookEynard} in order to complete the proof, but this has not been done so far.

In \cite{IwakiMarchal} a new method was found, for rank 2 systems, proving condition 4 for WKB-Lax systems, not relying on an insertion operator, but only relying on loop equations. The generalization of this method to higher dimensional representations is still missing.


Let us also mention results obtained from the opposite end: assuming only topological recursion, we get loop equations and Topological Type property, and the goal is to prove that we get a differential system. In other words, starting from topological recursion, one builds correlators $\hat W_{g,n}$, then define formal series $\hat W_n=\sum_{g=0}^\infty \hbar^{2g-2+n} \hat W_{g,n}$, and prove (in certain cases), that these lead to a formal differential equation $\hbar d\Psi=\Phi\Psi$, called the ``quantum curve". This method initiated in \cite{BE09} for the case of the Airy function and was successfully applied to other cases in \cite{DM14,MS12,N15}.

\subsection{Topological recursion}

It is proved in \cite{BEO13,EOFg}, that if a family of $\hat W_{n}$s satisfy the Topological Type property and satisfy loop equations, then they satisfy the topological recursion. 
The challenge for a given $\Phi(x,\hbar)$, is thus to prove the Topological Type property. The Topological Type property has already been proven for a variety of systems: the six Painlev\'{e} systems in \cite{IwakiMarchal}, and the $(p,2)$ minimal models in \cite{BookEynard}, plus incomplete proofs for $(p,q)$ minimal models in \cite{BBE14}.

We plan in a forthcoming article to prove it for all integrable systems whose spectral curve is a compact curve of genus zero, and satisfying a Lax equation.

%
%
%

\section{Conclusion}

We have generalized the derivation of loop equations of \cite{BE09,BBE14}, in a much more algebraic way. In particular our method does not use any ``insertion operator".
Another advantage of this new derivation is that it extends to all reductive Lie algebras, all Riemann surfaces and all choices of prime forms thus making it a general tool to be used in many different applications.

\section*{Acknowledgments}

The authors thank S. Ribault for fruitful discussions and helping develop the present method. B.E. thanks Centre de Recherches Math\'ematiques de Montr\'eal, the FQRNT grant from the Qu\'ebec government, Piotr Su\l kowski and the ERC starting grant Fields-Knots. O.M. would like to thank Universit\'e Lyon $1$, Universit\'e Jean Monnet and Institut Camille Jordan for financial support. The authors would also like to thank the organizers of the ``moduli spaces, integrable systems, and topological recursions'' workshop in Montr\'eal where part of this work was realized.

\end{document}